# A compact multi-planet system around a bright nearby star from the Dispersed Matter Planet Project

D. Staab[1,2], C.A. Haswell[1], J.R. Barnes[1], G. Anglada-Escudé[3], L. Fossati[4], J.P.J. Doherty[1,] J. Cooper[1], J.S. Jenkins[5], M.R. Díaz[5], M.G. Soto[3,5]

**The Dispersed Matter Planet Project targets stars with anomalously low Ca II H&K chromospheric emission. High precision, high cadence radial velocity measurements of the F8V star HD 38677 / DMPP-1 reveal four short period planets. DMPP-1 has log($R'_{HK}$) = -5.16 which** probably indicates **the presence of circumstellar absorbing gas arising from an ablating hot planet. The planets have $P_{orb}$ ~ 2.9 - 19 d, i.e., a compact planetary system with super-Earth (~3 $M_\oplus$) to Neptune-mass (~24 $M_\oplus$) planets. These irradiated planets may be chthonian: remnant cores of giant planets after mass-loss while crossing the Neptune desert. Modelling the possible long-term activity indicators while searching for Keplerian signals modifies the recovered planetary signals.** *A priori* **inferences about the presence of short period planets allowed the efficient discovery of the DMPP-1 planets. They have great potential for novel and informative follow-up characterisation studies.**

The Dispersed Matter Planet Project (DMPP) is a planet search motivated by the observation that close-orbiting planets often suffer mass loss, which may produce a dispersed circumstellar shroud of diffuse gas through which we view the host star[1-5]. Evidence for planetary mass loss was first found for giant planets, but there are dramatic examples of the phenomenon for low mass planets[6]. The Ca II H&K resonance lines provide a sensitive probe for the presence of diffuse circumstellar gas, so we used archival data to identify nearby, bright stars with depressed Ca II H&K line cores[7,8]. We anticipated that these stars might include nearby analogues of the Kepler compact multi-planet systems, and possible analogues and progenitors of Kepler 1520b. This paper presents DMPP-1, our first firm planetary system detection; it is a compact multi-planet system orbiting a V= 7.98 F8V star, HD 38677.

## Results
### Target selection
Stellar chromospheric activity is frequently parameterised in terms of log($R'_{HK}$), which conveniently places FGK stars on a scale where their activity levels can be easily compared. A summary of the definition and measurement of log($R'_{HK}$), and an illustration the photospheric and chromospheric fluxes as a function of spectral type can be found in[7,9]. A particularly useful property of log($R'_{HK}$) is that a completely inactive FGK main sequence star, corresponding to the quiet Sun devoid of active regions, has log($R'_{HK}$) = −5.1, independent of the star's effective temperature. For subgiant stars the basal level of observed chromospheric flux is dependent on effective temperature, $T_{eff}$[7,10]. DMPP therefore targets unevolved FGK stars exhibiting log($R'_{HK}$) < −5.1, i.e. main sequence stars with anomalously low chromospheric emission[7].

DMPP-1 was selected as a target due to its median activity level of log($R'_{HK}$) = -5.16 from six observations[10]. Only 4 previous radial velocity (RV) measurements with typical precision of 5 m s$^{-1}$ and over a baseline of 120 d were made by the Magellan Planet Search[11] with the MIKE

[1] School of Physical Sciences, The Open University, Walton Hall, MK7 6AA Milton Keynes, United Kingdom
[2] AVS, Rutherford Appleton Laboratory, Harwell, Oxford, OX11 0QX United Kingdom
[3] School of Physics and Astronomy, Queen Mary University of London, 327 Mile End Rd, E1 4NS London, United Kingdom
[4] Space Research Institute, Austrian Academy of Sciences, Schmiedlstrasse 6, A-8042 Graz, Austria
[5] Departamento de Astronomía, Universidad de Chile, Camino del Observatorio 1515, Las Condes, Santiago. Chile

spectrograph at the Las Campanas Observatory[12]. DMPP-1's properties are given in Table 1 (See Methods).

Radial velocity periodicities

We collected 148 high cadence HARPS[13] RV observations over a baseline of 760 d between 2016 Dec and 2018 Jan as shown in Fig. 1. Radial velocities were extracted[14] and SNR correlated systematics removed (see Methods for further details). The RV timeseries (Fig. 1) shows 15.5 m s$^{-1}$ peak-to-peak variability and obviously contains multiple signals (the expected RV jitter, 2.1 - 2.6 m s$^{-1}$, for DMPP-1 is much smaller[10]). The likelihood periodograms[15,16] in Fig. 2 indicate evidence for 4 Keplerian signals in the time series . We optimised the maximum likelihood parameters for each signal and then recursively searched for additional significant signals, resulting in the successive panels of Fig. 2. During this search, the model assumes eccentricity, $e = 0$.

The strongest peak of the first periodogram at $P = 18.57$ d (false alarm probability, FAP = $1.0 \times 10^{-26}$; $\Delta \log L = 71.9$) has a neighbouring peak at $P = 19.57$ d with $\Delta \log L = 70.5$ (see Methods) making it only 4.2 times less likely; thus there is some uncertainty in which of these aliases is the true period. Similarly, a $P = 17.78$ d signal is present with $\Delta \log L = 67.2$, making it 108 times less likely. The other peaks at $P = 40.7$ d, 51.2 d and 9.75 d are, respectively, over $9.9 \times 10^4$, $1.6 \times 10^4$ and $7.8 \times 10^6$ times less likely.

Adding a second Keplerian to the model yields a peak with $P_2 = 6.584$ d (FAP = $9.4 \times 10^{-17}$; $\Delta \log L = 49.1$) that has a sidelobe with almost equal likelihood at $P = 10.57$ d ($\Delta \log L = 49.0$). A third significant Keplerian with $P_3 = 2.882$ d (FAP = $1.6 \times 10^{-13}$; $\Delta \log L = 42.1$) is also found, which is distinct from its nearby sidelobe at $P = 3.146$ d by $\Delta \log L = 4.7$ (i.e. 113 times more likely). A further fourth significant Keplerian with $P_4 = 5.516$ d (FAP = $6.1 \times 10^{-4}$; $\Delta \log L = 18.4$) is present. Hereafter, we refer to these candidate signals as DMPP-1b, c, d & e. Table 2 (a) lists the maximum likelihood and false alarm probabilities for each signal. Forcing the $P_2 = 10.57$ d alias results in subsequent signals of $P_2 = 3.143$ d and $P_3 = 2.266$ d (these periods are identical or similar to the results that include the Ca II H & K S-index activity correlation below). The model fits with four Keplerians determined from the likelihood period search is shown in Fig. 1 (a), while Fig. 3 shows the folded data for each planet signal in turn with the other signals subtracted. The RV residuals have RMS values of 2.2, 1.6, 1.2 and 1.1 m s$^{-1}$ respectively for Keplerian solutions with 1, 2, 3 and 4 planets. No further signals with a false alarm probability below 40% ($\Delta \log L > 11.3$) were found.

In Table 2 (a), we list the maximum *a posteriori* parameters and probability increase, $\Delta \log \mathrm{MAP}$, after adding successive signals. The 68% confidence intervals for each parameter are estimated via Markov Chain Monte Carlo (MCMC) algorithms[17] (see Methods). Similarly, we derive estimates of minimum mass, $M_\mathrm{p} \sin i$, and semi-major axis, $a$, for each candidate planet. Our solution with 4 Keplerians corresponds to a Neptune mass planet in a warm orbit, with $M_\mathrm{p} \sin i = 24.35$ M$_\oplus$ and $a = 0.1462$ AU and 3 interior hot super-Earth planets with $M_\mathrm{p} \sin i = 9.61$ M$_\oplus$, 3.38 M$_\oplus$ and 4.36 M$_\oplus$ and respectively $a = 0.0733$, 0.0422 and 0.0651 AU.

Stellar Signals

Stellar activity can introduce spurious pseudo-periodic RV signals. We therefore examined the data for the indicators of line profile variability caused by stellar rotation or magnetic activity. We found the line bisector shape is stable, and the RV signals we detect are clearly dominated by parallel shift consistent with reflex orbital motion due to the presence of planets. We found some evidence of periodic variability in the line profile properties generally used as proxies of stellar activity, and we thoroughly investigated the correlations of these with the RVs (see Methods for details). Note, encouragingly, the most commonly used activity indicator, BIS, shows no periodic variability (Supplementary Fig. 2). While the RVs measured in individual seasons showed no correlations with line profile parameters, we found some evidence of long-term RV correlations with FWHM and S-values. Furthermore, there were some similarities between the activity proxy periodicities and those of the RVs, especially at longer periods (cf. Fig. 2 and Supplementary Fig. 2). This motivated further checks.

RV models with FWHM and S-index activity correlations
We carried out recursive likelihood periodogram analysis for the DMPP-1 RVs while incorporating either the FWHM or the Mount Wilson S-index (a measure of Ca II H & K core emission[18]) into the likelihood model via a linear activity correlation term. The activity time series correlation term is thus modelled simultaneously during the signal search process[15,16].

The global best fit including an RV-FWHM correlation results in a four Keplerian solution with identical or near-identical periods to those given in Table 2 (a); the $\Delta \log L$ value of the first Keplerian is *reduced* in significance by a factor of 67, with a FAP = $9 \times 10^{-25}$. The corresponding maximum *a posteriori* orbital periods are 18.56 d, 6.584 d, 2.882 d and 5.515 d, with respective $\Delta \log$ MAP values of 76.7, 53.6, 44.9 and 13.5.

Using instead a RV–S-index correlation yields periods and significances with some marked differences to those in Table 2 (a). Firstly, we find four significant periodicities with modified periods and a possible fifth periodicity with moderate significance. The periodograms are shown in Fig. 4 and the $\Delta \log$ MAP orbital periods are tabulated in Table 2 (b). The folded RV curves are shown in Fig. 5. $P_1'$ now favours the 19.6 d alias with a reduced amplitude, since the correlation term has removed some RV variability previously ascribed to the $P_1$ Keplerian. The second signal , $P_2'$ = 34.8 d, is the highest window function peak at < 190 d, and results from ill-constrained long-term variability. Subsequent recursively added Keplerian periodicities are close to, or aliases of the signals found in Table 2 (a).

Long timescale S-index variability is ubiquitous in main sequence FGK stars, and our 148 measurements are too few to disentangle a correlation with S-index which may be due to stellar rotation, a possible long-term low amplitude stellar activity cycle, and the four or five Keplerian signals. Our analysis with correlation terms has not revealed any compelling evidence that any of the RV periodicities in Table 2 (a) are caused by stellar activity; the analysis returns the same periods as detected initially, or alias solutions thereof. In particular the DMPP-1b signal we suspected might prove to be due to rotation persists for decorrelations with a variety of activity indicators (see Methods for details). Consequently, for simplicity, we adopt the solution reported in Table 2 (a) and Figs 1 (a), 2 and 3 as our preferred description of the Keplerian modulations detected. The 5.5d DMPP-1e signal no longer passes the 0.1% FAP threshold in the solution with correlation terms reported in Table 2 (b), so it does not appear there and could be considered marginal. It is clearly the least significant of our four planet detections, but it passes all the commonly adopted criteria for a significant detection, so we retain it in our preferred solution, Table 2(a) and Fig. 3

The stellar rotation period, and the first Keplerian signal
DMPP-1 clearly shows multiple periodic RV signals. Intriguingly, the $v \sin i$ and $R_*$ values (Table 1) imply a maximum stellar rotation period $P_{rot} / \sin i$ = 18.67 ± 1.465 d. This is very close to the period of DMPP-1b. The phase stability of the RV signal (Fig. 3) and the lack of significant correlations between RVs and the line bisectors or FWHM suggest a genuine planetary signal. If we attribute the $P_1$ = 18.57 d peak (Table 2 a) to stellar rotation we can constrain the DMPP-1 rotation axis to $i > 67°$. Our selection criteria favour edge-on orbital inclinations, and we would expect a compact multiplanet system to be co-planar and aligned with the stellar spin.

The (near-)synchronisation of the *maximum* stellar rotation and the period of the dominant planetary RV signal may be coincidence, or could be causally related. If we speculate that star-planet interactions (SPI) play a role we might expect peak chromospheric activity at inferior conjunction of the interacting planet. In fact, S-index minimum is at inferior conjunction of DMPP-1b, which might prompt us to imagine instead that the correlation between RVs and S-value arises from an over-density in the absorbing circumstellar material co-orbiting with DMPP-1b. The data are insufficient to support more than speculation.

Stability of the orbital configuration

The RVs of DMPP-1 indicate a compact system of planets. In particular, DMPP-1c and DMPP-1e are very close neighbours (Table 2 (a)), with semi-major axis difference of only $\Delta a = 0.0082$ AU. Assuming $e = 0$, the Hill radius of DMPP-1c is

$$r_H = 0.0733 \text{ AU } \sqrt[3]{\frac{9.61 \, M_\oplus}{3 \times 1.21 M_\odot}} = 0.00146 \text{ AU}.$$

Within the uncertainties in $a$, the closest approach of DMPP- e is 4.8 $r_H$.

We performed numerical simulations to examine the orbital stability of the solutions in Table 2 (a). We used the IAS15 integrator[19] within the N-body orbital integrator, REBOUND[20], since we do not know *a priori* whether close encounters occur. We considered two cases: firstly starting with circular orbits ($e = 0$) for all planets, and secondly for elliptical orbits with the range of upper limits to the eccentricity listed in Table 2 (a). We integrated for $10^6$ yrs, which corresponds to $1.27 \times 10^8$ orbits of the inner planet, DMPP-1d. Fig. 6 shows the evolution of semi-major axis, $a$, and eccentricity, $e$, with time for the two cases. The long-term variation in eccentricity repeats on a timescale of 1450 years and there are lower amplitude variabilities on timescales of up to a few tens of years, depending on the starting eccentricity. These shorter periodic variabilities are most pronounced for DMPP-1c and DMPP-1e as expected. Initially circular orbits (Fig. 6a) maintain small eccentricities throughout. In all simulations, the orbits remain stable.

Investigations revealed the orbits of DMPP-1c and DMPP-1e (Table 2 (a)) become unstable only when both planet masses are multiplied by a factor of greater than 9 (starting with $e = 0$) or greater than 8 (starting with $e > 0$). Similarly, multiplying the mass of DMPP-1b by $> 12$ times results in instability. We are thus confident that the solution in Table 2(a) describes a stable planetary system.

Likelihood of Transits

If we assume a completely random orbital orientation, the probability of transit for DMPP-1d is 14%. However, we selected DMPP-1 as a target[7,9] on the hypothesis that our line of sight to the chromospherically active regions of the star is filled with absorbing circumstellar gas. If this gas is ablated from the close-orbiting planets, then angular momentum considerations suggest that we are viewing the system approximately edge-on. The very planet discoveries we report suggest that the fundamental hypothesis of DMPP is correct. Haswell et al. (Dispersed Matter overview article, this issue) present a statistical analysis implying a less than 0.1% probability that DMPP's first three planetary system discoveries would have been made with randomly selected targets. Consequently, we expect a significantly higher than random transit probability for the DMPP-1 planets.

DMPP-1 is a bright, nearby star. It would be amongst the brightest dozen known host stars of transiting planets, with apparent magnitude V=8 or brighter, if it does exhibit transits. The least massive of these transiting planets is HD 219134b with a mass of 3.8 $M_\oplus$. DMPP-1d is less massive and offers the potential for high quality characterisation of an exoplanet of about three Earth masses. Furthermore, it is possible that DMPP-1 harbours further planets which produce reflex RV signals below our present detection threshold. DMPP-1 is a nearby analogue of the Kepler compact multi-planet systems. It is thus an important system for continued radial velocity observations as it offers the opportunity to observe the dynamical interactions between the planets, particularly if transits occur. These tightly-packed systems provide particularly valuable empirical tests of ideas about how planetary systems form and evolve[21-24]. Some of the masses produced by the transit timing variation method for the Kepler systems are surprising[25] and these systems are too distant and hence faint for high precision RV measurements. DMPP-1 provides a valuable bench-marking opportunity. With DMPP-1 there are still only 9 compact systems with three or more planets and V< 8, making them amenable to detailed follow-up observations. Among these 9, DMPP-1 is among the most compact, hosts among the lowest mass planets and has the hottest host star (Supplementary Fig. 4). These characteristics are consistent with the presence of ablated planetary material forming a detectable circumstellar gas shroud. Furthermore, these properties suggest detailed study of the DMPP-1 system may be particularly informative for the evolutionary mechanisms sculpting the Neptunian desert. It is consequently a very important system for space-based photometry with TESS and CHEOPS.

Compositional Studies

Irrespective of whether the planets transit, our line of sight to the star is filled with circumstellar gas. Any azimuthal inhomogeneities in this (presumably orbiting) gas present the opportunity to apply transmission spectroscopy techniques. By comparing spectra viewed through the highest column density of circumstellar gas with those viewed through lower column density, we can detect the presence of species with varying column density. This has a significant advantage over conventional transmission spectroscopy because the gas absorbs over the entire stellar disc, rather than being confined to the small atmospheric annulus surrounding a transiting planet. Of course, this is coupled with the disadvantage that we probably never have a line of sight completely devoid of circumstellar gas, and the compositional signal will be directly proportional to the variation in the column density of each species under examination, assuming we are on the linear part of the curve of growth.

For a compact system of low mass planets, these compositional studies are particularly interesting. Planets as close to their star as the ones we have detected are unlikely to retain their atmosphere. The absorbing gas could originate from liquid magma, which on Earth is typically 1100 – 1500 K[26]. The equilibrium temperatures $T_{eq}$ in Table 2 suggest that all our planet candidates except DMPP-1b could potentially have liquid magma at the surface. Alternatively, even with solid surfaces, it is possible that the proximity of the planets to the host star will result in sputtering from the stellar wind particles, leading to relatively dense atmospheres that are rich in refactory elements[27]. It is also possible that DMPP-1 hosts an as yet undetected catastrophically disintegrating low mass rocky object analogous to Kepler 1520b. The latter has a 16 hour orbit and its inferred mass is ~0.1 $M_\oplus$, well below our RV detection threshold. Examination of the circumstellar gas composition thus may reveal the minerology of the ablating rocky surfaces. If the planets transit, the system offers exciting prospects to directly probe the mass-radius-composition relationship.

The Neptunian Desert

The innermost planet candidates, DMPP-1c,d,e with super-Earth to sub-Neptune masses and periods of 2.9 d – 6.6 d lie below the boundary of the Neptunian desert[28]. It is possible that these bodies are chthonian planets: the remnants of gas giant planets that have migrated inwards and subsequently lost mass through stellar insolation or Roche lobe overflow. In this scenario, planets move downwards in the period-mass plane to join the more densely populated low-mass (or radius) short-period population of planets[28]. The mass-loss timescale will become dramatically longer once the gaseous envelope is completely ablated, hence the population density of rocky objects below the Neptune desert exceeds that of larger planets crossing the desert.

Star Planet Interactions

The most massive planet we detected, DMPP-1b, has an orbital period suspiciously close to the rotation period we infer for the star. The stability of the Keplerian signal over the long baseline of our observations along with the activity metrics imply the signal is genuine orbital motion. It is possible that SPI causes the similarity of the inferred stellar rotation period and DMPP-1b's orbital period. The obvious next steps are (i) to search for $P_{rot}$ photometrically (ii) to extend the RV baseline, thus further assessing the signal's stability and definitively establishing the correct ephemerides for this system.

## Summary and Concluding Remarks

With a carefully targeted RV campaign comprising 148 observations between 2015 December and 2018 January, we find a compact system of Neptune and super-Earth mass planets orbiting DMPP-1. Although we favour a solution without a stellar activity-correlation, we note that further observations are needed to fully disentangle aliases and characterise a possible stellar contribution. In the most conservative interpretation, if we attribute DMPP-1b purely to stellar rotation, we discovered a system of 3 low mass planets with only 148 RV measurements. In fact we think the four planet solution is more likely because the DMPP-1b signal is highly coherent and persists even when we include RV correlation with activity indicators in our period search. Our observations demonstrate an excitingly efficient use of telescope time, particularly as the DMPP-1 planets have great potential for novel and informative follow-up characterisation studies.

**Table 1**. HD 38677 stellar parameters

| Parameter | Value | Notes |
|---|---|---|
| Spectral Type | F8V | [14] |
| Parallax [mas] | 16.03 ± 0.25 | [15] |
| Distance [pc] | 62 ± 1 | [15] |
| $V$ [apparent mag.] | 7.98 | [15] |
| $B$-$V$ [apparent mag.] | 0.581 ± 0.013 | [15] |
| $\log(R'_{HK})$ | -5.16 | [11] |
| $T_{eff}$ [K] | 6196 ± 29 | This work |
| $\log g$ [cm s$^{-2}$] | 4.41 ± 0.21 | This work |
| $v \sin i$ [kms$^{-1}$] | 3.41 ± 0.24 | This work |
| $v_{mac}$ [kms$^{-1}$] | 1.50 ± 0.14 | This work |
| $R_*$ [R$_\odot$] | 1.26 ± 0.02 | This work |
| $M_*$ [M$_\odot$] | 1.21 ± 0.03 | This work |
| Age [Gyr] | 2.01 ± 0.54 | This work |

Table 2: (a) Best fit Keplerian parameters with no correlations and (b) with S-index correlation included – parameters and planet candidates are denoted with a prime (′) to distinguish them from the preferred solutions in (a). We give the false alarm probability for the peak $\Delta \log L$ indicated in Figs 2 & 4. $\Delta \log L$ quantifies the improvement to the likelihood due the introduction of each Keplerian; similarly, $\Delta \log$ MAP reports the log maximum *a posteriori* probability. We report the maximum *a posteriori* values of the period, $P$, velocity amplitude, $K$, eccentricity, $e$, mean longitude at reference epoch $t_0$ (the first observation), $\lambda = M_0 + \omega$ (where $M_0$ is the mean anomaly at $t_0$) and the longitude at periastron, $\omega$. The sum $M_0 + \omega$ is better constrained than either individual quantity when $e$ is small. Uncertainty ranges in parentheses are from the MCMC posterior samples. Significant period aliases found in the log-likelihood periodograms in Fig. 2 along with relative significances are indicted in italics in square parentheses. The RV offset $\gamma_{HARPS}$ is relative to the HARPS-TERRA template and $\sigma_{HARPS}$ is the excess white noise. The planet parameters assume a stellar mass, $M_* = 1.21 \pm 0.03$ (see Table 1). The final 4 rows of each sub-table list the minimum planet mass, $M_p \sin i$, Semi-major-axis, $a$, the maximum equilibrium temperature of the planet, $T_{eq}$, for a perfect blackbody (BB) and for a planet with an atmosphere (Atmos) following [46] (see Methods).

| | (a) PREFERRED SOLUTION Keplerians (RVs only / no additional correlations) | | | |
|---|---|---|---|---|
| Parameter | DMPP-1b | DMPP-1c | DMPP-1d | DMPP-1e |
| FAP | $1.0 \times 10^{-26}$ | $9.4 \times 10^{-17}$ | $1.6 \times 10^{-13}$ | $6.1 \times 10^{-4}$ |
| $\Delta \log L$ | 71.9 | 49.1 | 42.1 | 18.4 |
| $\Delta \log$ MAP | 80.9 | 53.8 | 44.3 | 14.5 |
| $P$ [d] | 18.57 (18.56 - 18.58) | 6.584 (6.582 - 6.587) | 2.882 (2.881 - 2.883) | 5.516 (5.512 - 5.518) |
| | [*19.57 d - 4.2x less likely*] | [*10.57 d - almost as likely*] | [*3.146 d - 113x less likely*] | |
| $K$ [ms$^{-1}$] | 5.162 (4.853 - 5.394) | 2.884 (2.414 – 3.044) | 1.326 (1.196 - 1.475) | 1.316 (0.958 - 1.522) |
| $e$ | < 0.083 | < 0.057 | < 0.070 | < 0.070 |
| $\lambda = M_0 + \omega_0$ [deg] | 262.0 (258.2 - 267.3) | 292.3 (284.2 – 300.4) | 47.8 (39.3 - 61.0) | 319.6 (294.7 – 326.2) |
| $\gamma_{HARPS}$ [ms$^{-1}$] | 1.88 (1.69 – 2.06) | | | |
| $\sigma_{HARPS}$ [ms$^{-1}$] | 0.59 (0.47 - 0.71) | | | |
| $N_{obs}$ | 148 | | | |
| Data baseline [d] | 763.2 (2.1 years) | | | |
| $t_0$=BJD-2400000 [d] | 57375.58981035 | | | |
| $M_p \sin i$ [$M_E$] | **24.27** (22.68 - 25.43) | **9.60** (8.02 - 10.13) | **3.35** (3.01 - 3.73) | **4.13** (2.99 - 4.79) |
| $a$ [AU] | 0.1462 (0.1450 - 0.1474) | 0.0733 (0.0726 - 0.0739) | 0.0422 (0.0419 - 0.0426) | 0.0651 (0.0646 - 0.0656) |
| $T_{eq}$ (BB) [K] | 877 | 1239 | 1632 | 1314 |
| $T_{eq}$ (Atmos) [K] | 816 | 1153 | 1519 | 1223 |

| | (b) Keplerians with S-index correlation | | | | |
|---|---|---|---|---|---|
| Parameter | DMPP-1b′ | DMPP-1c′ | DMPP-1d′ | DMPP-1e′ | DMPP-1f′ |
| FAP′ | $3.9 \times 10^{-24}$ | $7.7 \times 10^{-15}$ | $1.1 \times 10^{-6}$ | $2.8 \times 10^{-9}$ | $2.9 \times 10^{-2}$ |
| $\Delta \log L'$ | 66.2 | 44.7 | 25.2 | 32.4 | 14.5 |
| $\Delta \log$ MAP′ | 79.3 | 38.7 | 28.5 | 36.8 | 17.3 |
| $P'$ [d] | 19.61 (19.60 - 19.66) | 34.74 (34.37 – 34.96) | 3.144 (3.142 - 3.145) | 6.454 (6.444 – 6.457) | 2.281 (2.278 - 2.281) |
| $K'$ [ms$^{-1}$] | 4.161 (3.999 - 4.808) | 3.472 (2.100 - 3.907) | 1.524 (1.268 - 1.591) | 1.498 (1.283 – 1.625) | 0.778 (0.607 - 0.926) |
| $e'$ | < 0.064 | < 0.070 | < 0.079 | < 0.069 | < 0.072 |
| $\lambda' = M_0 + \omega_0$ | 289.6 (287.0 – 306.0) | 52.8 (52.5 – 78.5) | 60.4 (48.1 - 716.8) | 272.4 (238.2 – 282.1) | 73.4 (22.7 - 96.6) |
| $\gamma_{HARPS}$ [ms$^{-1}$] | 0.05 (-0.13 - 0.67) | | | | |
| $\sigma_{HARPS}$ [ms$^{-1}$] | 0.43 (0.42 – 0.66) | | | | |
| $M_p \sin i'$ [$M_E$] | 19.92 (19.12 – 23.07) | 20.12 (17.90 - 22.74) | 3.95 (3.28 - 4.14) | 4.95 (4.23 – 5.36) | 1.82 (1.42 - 2.16) |
| $a'$ [AU] | 0.1517 (0.1505-0.1530) | 0.2220 (0.2206 – 0.2244) | 0.0447 (0.0444 - 0.0451) | 0.0723 (0.0717 - 0.0729) | 0.0361 (0.0358 - 0.0364) |
| $T_{eq}'$ (BB) [K] | 861 | 712 | 1586 | 1247 | 1765 |
| $T_{eq}'$ (Atmos) [K] | 801 | 662 | 1476 | 1161 | 1642 |

## Acknowledgements

This work is based on observations collected at the European Organisation for Astronomical Research in the Southern Hemisphere under ESO programmes 096.C-0876(A) and 098.C-0269(A), 098.C0499(A), 098.C0269(B), 099.C-0798(A) and 0100.C-0836(A). D.S. was supported by an STFC studentship. CAH and JRB were supported by STFC Consolidated Grants ST/L000776/1 and ST/P000584/1; DS was supported by an STFC studentship. GA-E was supported by STFC Consolidated Grant ST/P000592/1; JSJ acknowledges support by FONDECYT grant 1161218 and partial support from CONICYT project Basal AFB-170002. M.R.D. acknowledges the support of CONICYT-PFCHA/Doctorado Nacional-21140646, Chile and project Basal AFB-170002. The research leading to these results has received funding from the European Community's Seventh Framework Programme (FP7/2007-2013 and (FP7/2013-2016) under grant agreements number RG226604 and 312430 (OPTICON). These results were based on observations awarded by ESO, OPTICON and OHP using HARPS, HARPS-N and SOPHIE. This research has made use of the SIMBAD data base, operated at CDS, Strasbourg, France.

## Author Contribution

DS performed target selection, contributed to writing of proposals, initial RV analyses and technical details of the paper. CAH leads all aspects of the DMPP collaboration, secured the funding, wrote the proposals, and introductory material to the paper. JRB contributed to proposals, performed final RV analyses, wrote the initial paper draft, the technical sections and made the figures. GA-E provided software and expertise. LF contributed to the analysis and proposal writing. JSJ and MGS provided expertise on stellar activity, the log $R'_{HK}$ metric and contributed stellar parameter analyses. DS, CAH, JRB, JD, JC and MRD performed observations with HARPS. All authors were given the opportunity to review the results and comment on the manuscript.

## Author Information

Reprints and permissions information is available at www.nature.com/reprints. Correspondence and requests for materials should be addressed to Carole.Haswell@open.ac.uk.

**Figure 1:** RV measurements and Keplerian solutions corresponding to parameters tabulated in Table 2. The plotted RV uncertainties are derived by HARPS-TERRA[14] from the extracted HARPS-DRS spectra (see Methods). Weighted nightly averages are shown as single, darker and thicker points with corresponding error bars where appropriate. The upper panels in **(a)** show our preferred four-Keplerian signal where the maximum likelihood solution is plotted with a dashed curve and the maximum *a posteriori* (MAP) solution with a solid curve. The lower panels show the residuals after removing the MAP fit. The solutions with S-index correlation term included are shown in **(b)**. Here we show the MAP solutions for a four-Keplerian (dashed curve) and five-Keplerian (solid curve) solution; the second signal in both these cases is the 34.7 d periodicity linked to activity and/or window function. Residuals in the lower panels are after subtraction of the five-Keplerian solution (see text for details).

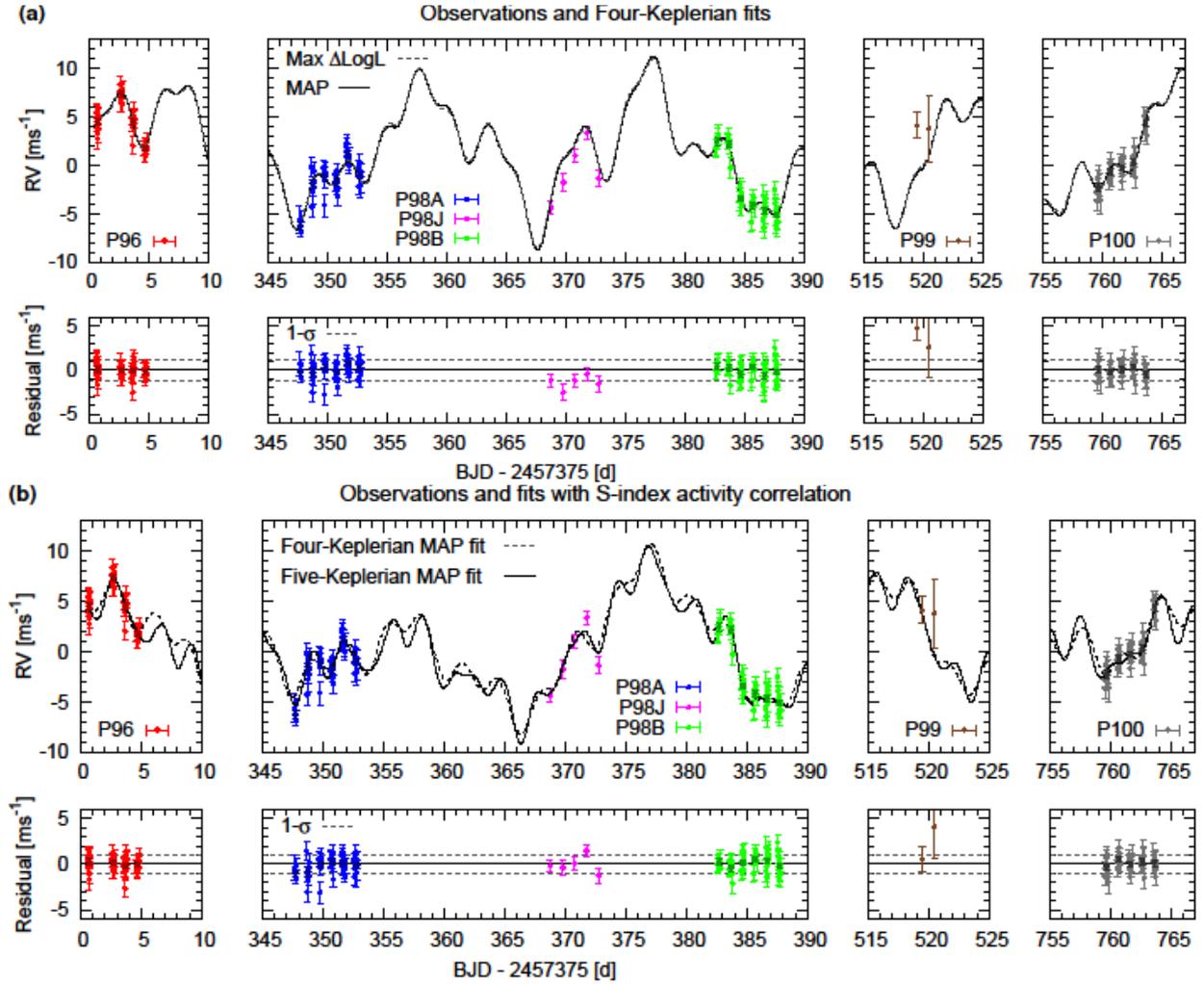

**Figure 2:** Log-likelihood (Δ log $L$) periodograms of the window function and the four-Keplerian search for the complete data set. The maximum likelihood and *a Posteriori* periods are identical at the indicated precision, both here and in Table 2, and are indicated by the vertical grey dashed lines. False alarm probability levels at 10%, 1.0%, 0.1% (bottom to top) are shown as horizontal dashed lines.

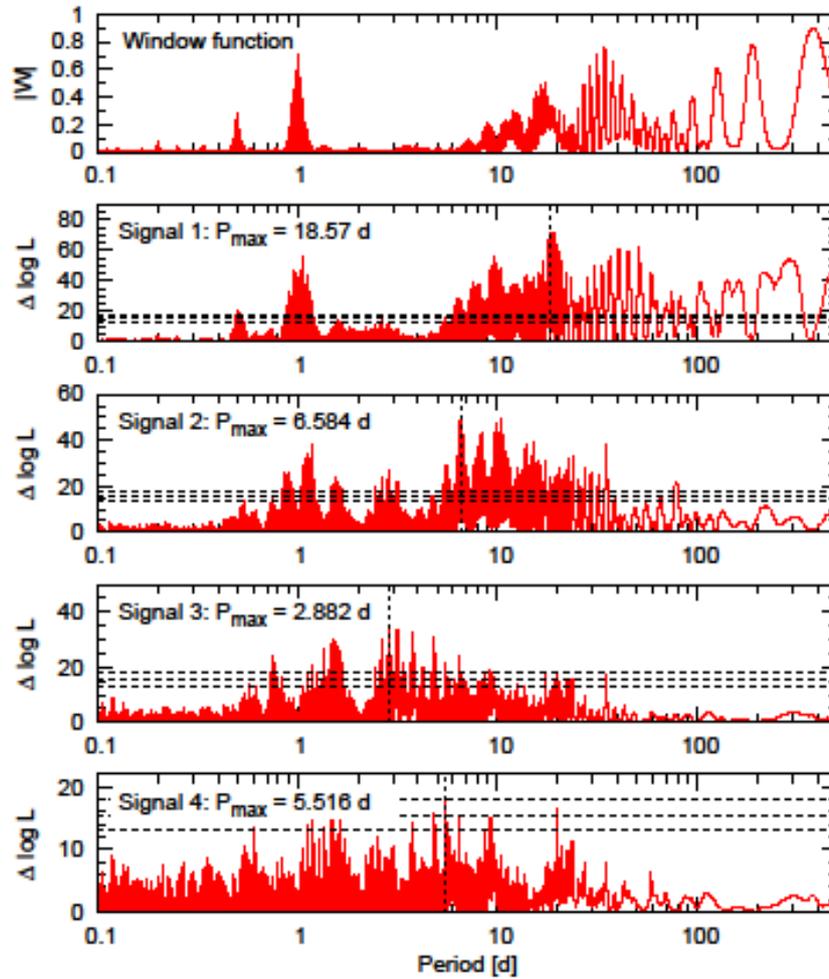

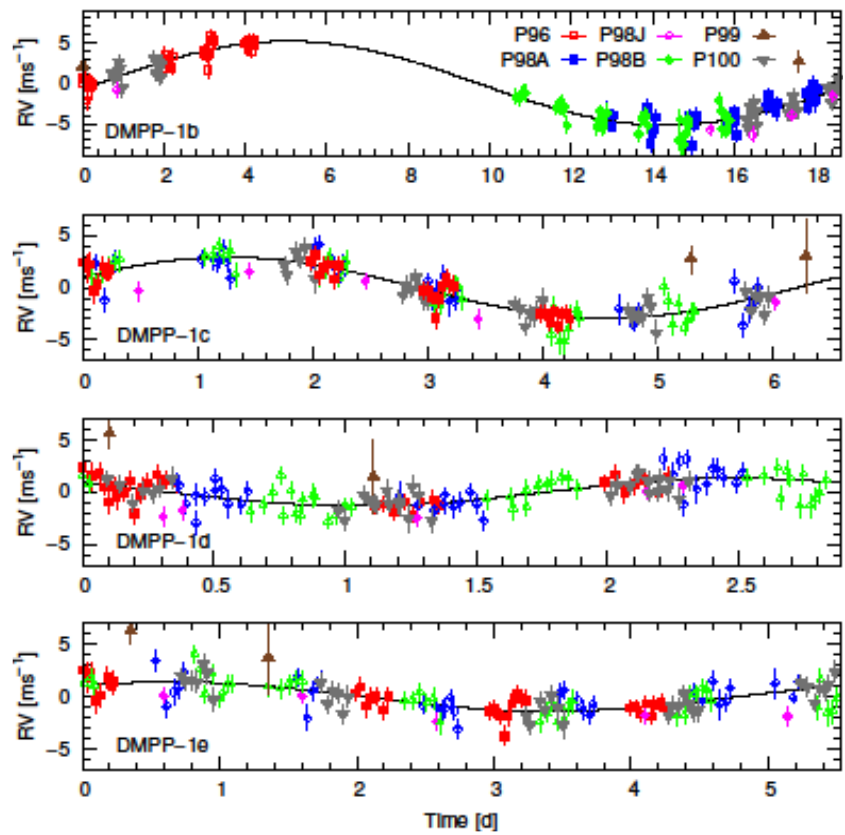

**Figure 3:** The phase folded RV data and corresponding uncertainties for each Keplerian signal (see Table 2(a)) after subtracting the other signals. See Fig. 1 for observations and combined signal.

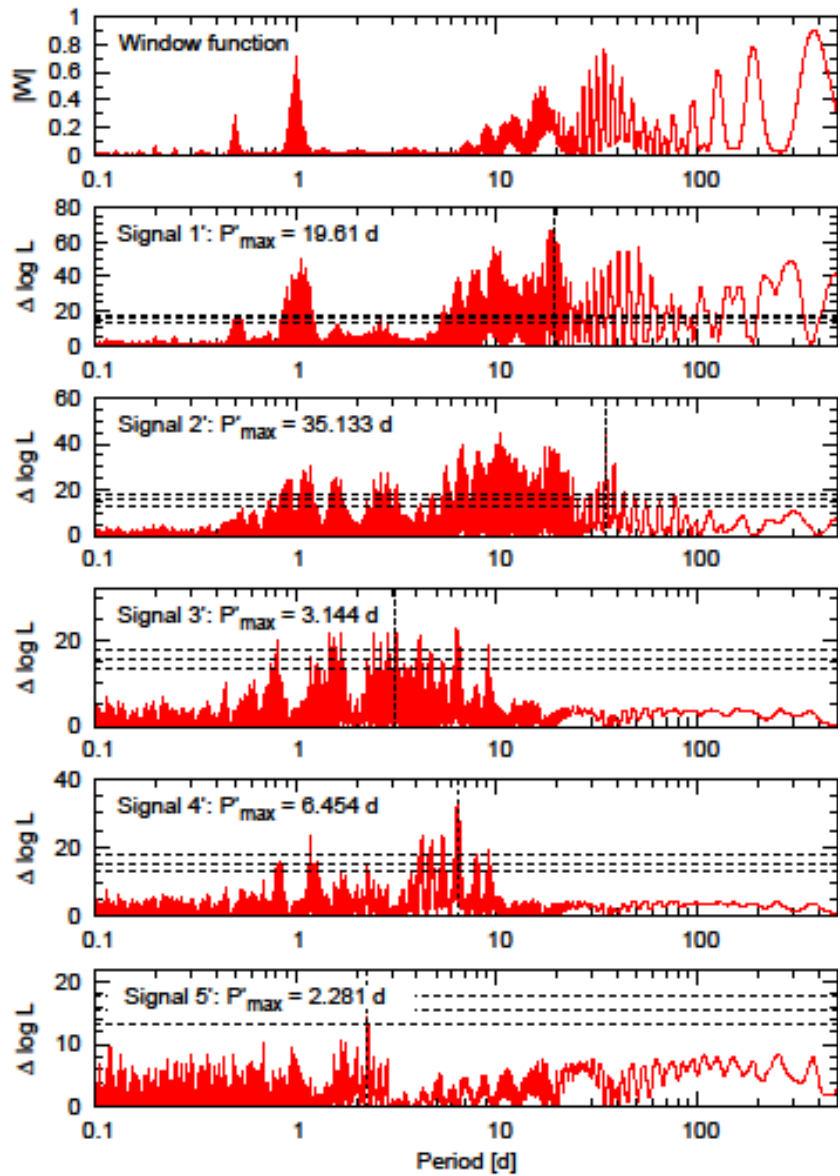

**Figure 4:** Log-likelihood periodograms as in Fig. 2 with inclusion of S-index activity correlation. Four significant periodicities are recovered. The fifth signal has marginal FAP = 0.029 but shows Δ log $L$ = 14.5 and Δ log MAP = 17.3 (see Table 2 (b)).

**Figure 5:** The phase folded RV data and corresponding uncertainties for each Keplerian signal from the five Keplerian solution that incorporates an S-index-RV activity correlation term (Table 2(b)). As in Fig. 3, each Keplerian signal is shown after subtracting the other signals.

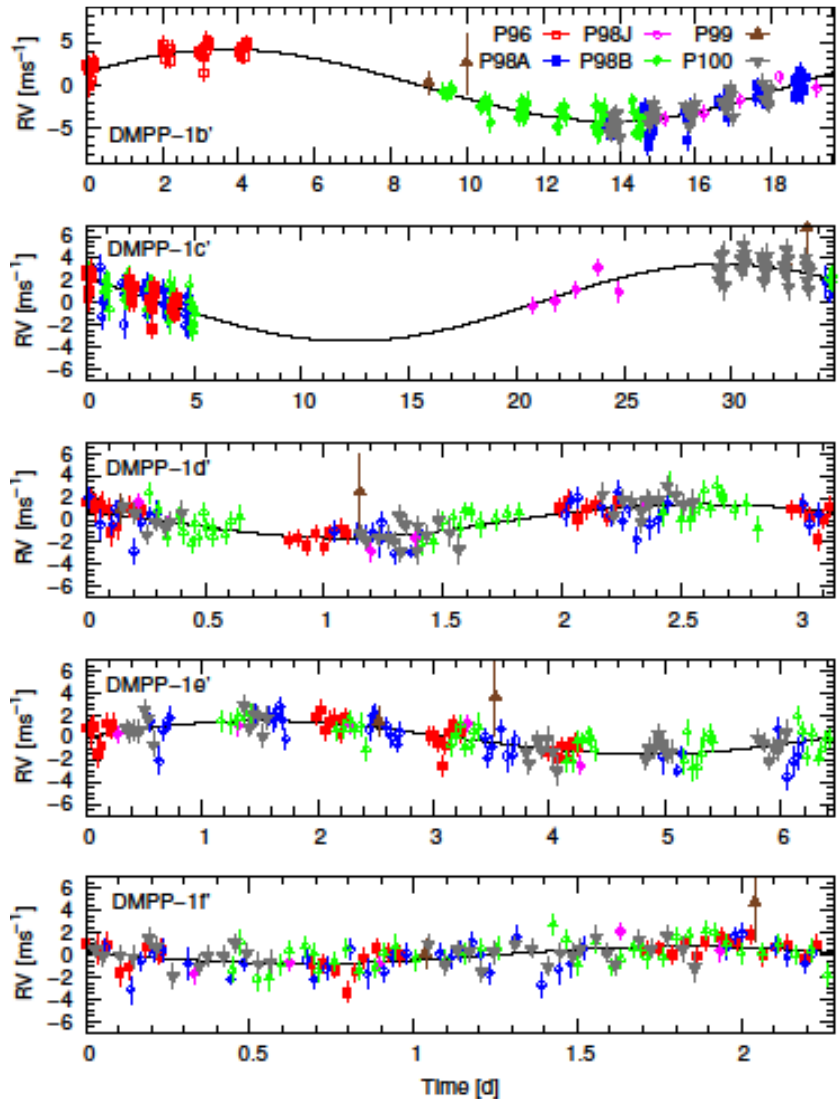

**Figure 6:** Numerical simulations of the preferred four-Keplerian solution for DMPP-1 (Table 2 (a)). Evolution of semi-major axis, a (left column), and eccentricity, e (right column); initially circular

orbits in the upper panels, with lower panels showing initially elliptical orbits using the upper limit e values from Table 2 (a). DMPP-1c and e show the largest variations in semi-major axis, a. the orbits are stable in all cases. The eccentricities invariably oscillate on a 1450 year timescale (shown), which repeats throughout the 106 year simulations.

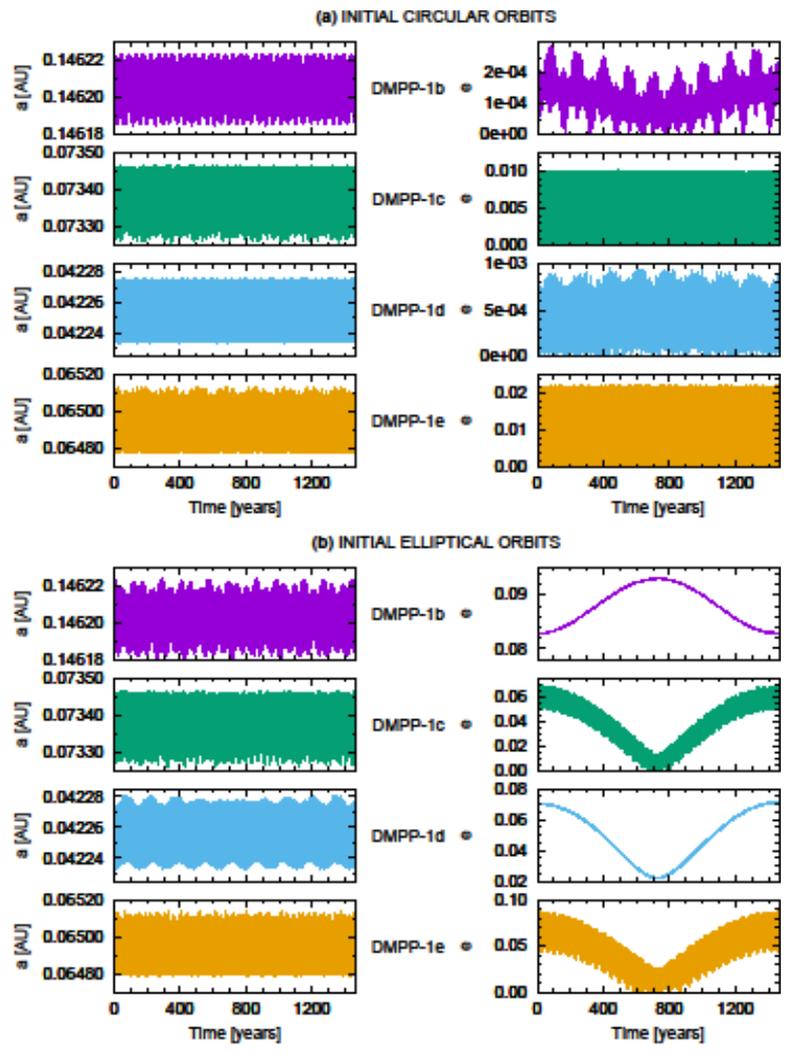

## Methods

### RV Observations

Observations were made with HARPS over several semesters (Fig. 1). In 2015 Dec (hereafter P96), observations of DMPP-1 were made at intervals ranging from 15 to 76 minutes, with a 50 minute average cycle time while the star was accessible on 4 nights out of 5 consecutive nights. RV variability on a timescale of > 1 d is present, spanning a range of 5.2 m s$^{-1}$. Two runs were allocated in P98: 2016 Nov/Dec (P98A) and 2017 Jan (P98B), and we made observations via a time-swap agreement (P98J). Two observations were taken in 2017 May (P99) under poor conditions. A 5 night run in 2018 Jan (P100) completes the observations reported herein. In total, 148 observations were made: 32 points in P96, 35 points in P98A, 5 points in P98J, 41 points in P98B, 2 points in P99B and 34 points in P100.

Exposure times were 900 seconds for all but two observations. Variable extinction and seeing led to a SNR range in the central order of 66 - 221 (mean and median of 157 and 160) and $\sigma_{RV}$ values 0.5 - 1.4 m s$^{-1}$ for all points except a single P99 observation with $\sigma_{RV}$ = 3.5 m s$^{-1}$. Mean and median uncertainties are respectively 0.91 m s$^{-1}$ and 0.86 m s$^{-1}$.

### Data Reduction

RVs were obtained using HARPS-TERRA[14] which uses least-squares minimisation of the difference between the individual target spectra and a high SNR template spectrum, derived from co-addition of target spectra. Extracted wavelength calibrated spectra from the standard HARPS Data Reduction Software (HARPS-DRS) output are used as the input spectra for HARPS-TERRA (see https://www.eso.org/sci/facilities/lasilla/instruments/harps/doc/DRS.pdf). HARPS-TERRA propagates errors to derive an uncertainty for each measurement. We corrected the RVs for the low amplitude SNR-correlated systematic effect due to charge transfer inefficiency[7] (see DMPP overview paper, this issue). Our DMPP-1 observations have signal-to-noise ratio in échelle order 60 of ~93 - 251, corresponding to CTI corrections of 0.21 and -0.81 m s$^{-1}$. We use bisector Span (BIS) and full width at half-maximum (FWHM) values from the HARPS-DRS.

### The S-index Activity Indicator

HARPS-TERRA produced values for the S-index, which is a parameterisation of the stellar activity-related emission in the Ca II H&K line cores. We also retrieved the raw spectra from the ESO archive and calculated S-values after taking particular care with the background subtraction[29]. In our subsequent analyses, we used our bespoke reduction of S-values.

### HD 38677 / DMPP-1 stellar parameters

Our code `SPECIES`[30] uses high resolution spectra and measures the equivalent widths (EWs) for a set of Fe I and Fe II lines. The temperature, metallicity, surface gravity and microturbulence velocity of the stellar photosphere were derived using the `MOOG` code[31], along with `ATLAS9` model atmospheres[32] and the EW values, by solving the radiative transfer equation assuming local thermodynamic equilibrium. Mass, radius and age were derived using the `ISOCHRONES`[33] package, which interpolates through a grid of `MIST` models[34]. Chemical abundances were derived by measuring the EW of a set of absorption lines for each element, and using the atmospheric parameters obtained previously. Finally, rotational and macroturbulence velocity are found by temperature relations and absorption line fitting with synthetic spectra.

### Frequentist Approach to Finding Periodic RV Modulations

We used a frequentist approach to planet recovery to search for periodicities in our data[15,16]. The software allows the eccentricity of all Keplerian signals to be set to zero, simplifying the model. It is relatively common practise to force circular orbits in RV analysis when recovering low amplitudes and/or short periods[35,36], especially when limited data is available. For the short period orbits of prime interest here, tidal circularisation is expected to significantly dampen orbital eccentricities over the systems' lifetime[37]. The likelihood fitting procedure also includes an optional linear trend term and activity correlation terms. This is useful when testing the effect of putative correlations and long-term slopes on Keplerian signal recovery. We plot $\Delta \log L$ periodograms, which measure the improvement in the natural logarithm of the likelihood function. The frequentist False Alarm Probability (FAP) of a planetary signal can then be computed from the $\Delta \log L$ value. This is done via the likelihood-ratio test[38,39] The FAP estimates the fraction of times a signal with a given $\Delta \log L$ would be seen in the data set purely due to unfortunate arrangement of Gaussian noise. There is no universally agreed value in the literature on the FAP threshold for claiming statistically significant planet detections[40]. Thresholds of 1% and 0.1% are typically used. Formally significant signals must also be well-constrained in period from above and below - for periods significantly longer than the baseline of observations this will not be the case. Such signals can be formally extremely significant, but nevertheless not constitute detections because a strict periodicity cannot be verified.

Posterior parameter determination and uncertainties

We obtain empirical parameter uncertainties for local maximum likelihood peaks from *a posteriori* samples, generated using Markov Chain Monte Carlo (MCMC) algorithms[17]. We assume uniform priors for all parameters except for the eccentricity, *e*, where we impose a semi-Gaussian prior with zero mean and width, $\sigma_e = 0.05$, thereby imposing a preference for low eccentricity orbits. By slowly cooling MCMC chains we converge to maximum *a posteriori* (MAP) solutions for each signal. The MAP period solutions either agree exactly with the maximum likelihood periods to the quoted precision in Table 2 (a) or differ by only one part in the last quoted decimal place. We then sample parameter space with MCMC simulations to obtain estimates of the parameter distributions.

## Data availability Statement

The data that support the plots within this paper and other findings of this study are available from the corresponding author upon reasonable request.

## Methods section references

## Supplementary Information

### Examination of the Line Bisectors

Supplementary Fig. 1 shows the line bisector shapes, which appear stable over the entire span of our observations. The few deviations from the median bisector shape are attributable to noise, with the low signal to noise measurement from P99 being by far the most deviant. The right-hand panel of Supplementary Fig. 1 shows the bisectors with the derived RVs subtracted, demonstrating the variations are dominated by parallel shifts in velocity, consistent with the presence of the inferred planets.

### Searches for Periodic line profile variability and correlations with RVs

We calculated the likelihood periodograms for the line bisector span (BIS), full width at half-maximum (FWHM) and Ca II H&K S-index to assess the possible contamination of RVs by low amplitude stellar line profile variations. The S-index parameterises the Ca II H&K line core emission in a simpler way than $\log(R'_{HK})$ and is suitable for examining temporal variations in a single star, but not intercomparisons between stars of differing effective temperatures. Supplementary Fig. 2 shows that there are no significant periodic signals present in the BIS time series, consistent with the lack of variation in the line bisectors shown in Supplementary Fig. 1.

By contrast, the S-index periodogram shows extremely significant signals with FAPs well below the 0.1 per cent FAP threshold. The pattern of alias peaks clearly matches that of the window function indicating an ill-determined long period signal. The S-index shows variability from season to season as is apparent in the relevant panel of Supplementary Fig. 3; e.g., the P98 S-index values are 0.0029 less on average than the P96 values. Adding a linear trend at the signal search level makes little difference to the periodogram. Peak $\Delta \log L$ power is seen at $P = 36.1$ d ($P = 36.3$ d with a linear trend included in the periodogram fits). This signal is very close to the 35.1 d peak seen in the second Keplerian (Figure 2, panel 3), and is most likely associated with the highest window function peak (below 187 d) at 34.5 d. However, the window function alone should not produce such significant peaks. Folding the S-index time series on the 36.1 d period suggests that the power in this period is dominated by the difference in mean S-index values, particularly in P98A and P98B compared with P100 (see Supplementary Fig. 3 caption). On folding the S-index time series, the one month separation of P98A and P98B is clearly responsible for the preferred 36.1 d period. Other peaks in the S-index periodogram, such as at P ~ 85.6 d result in incomplete phase coverage of only 45%. The shorter period peaks around P ~ 17.1 d are close to the potential maximum rotation period of DMPP-1, but may also be aliases of the longer period peaks. Without further observations, it is difficult to determine whether shorter period rotation modulated variability or longer term activity variability[1] is causing the apparent S-index variability. Both may be responsible. The FWHM time series shows signals at 1.6 d and 2.6 d close to 0.1 per cent FAP. Adding a linear trend to the likelihood model (blue line in Supplementary Fig. 2 Panel 3) removes longer period signals. There are no peaks with less than 0.1 per cent FAP and none that directly match the three RV signals.

We tested the significance of linear correlations between the proxies tracing potential stellar signals (FWHM, S-index and BIS) and the RVs. The trends are shown in Supplementary Fig. 3 while Supplementary Table 1 lists the correlation test values[2]. Where there were sufficient observations (i.e. P96, P98 and P100), we determined the correlation parameters on each data set in isolation; Supplementary Table 1 shows that there are no significant correlations. However, when the full data set is used, we find weak to moderate correlations between FWHM and RV and S-index and RV

(bottom row of Supplementary Table 1). The *r* and *p* statistics particularly indicate a significant moderate correlation between S-index and RV. The correlations and offsets observed in the FWHM and the S-index time series could be driven by a stellar activity cycle or the shorter stellar rotation related period[1]. The presence of long-term RV correlations with FWHM and S-values, and the similarity of the activity proxy periodicities and those of the RVs motivated us to explore RV solutions including correlation terms. Without a longer timespan of observations it is difficult to reliably assess the timescales on which the correlations are seen. For this reason we adopted the RV solution without activity.

Equilibrium Temperature calculations

The equilibrium temperatures for each planet listed in Table 1 (Main Article) are for a blackbody (BB) and for a planet with a dense atmosphere with a Bond albedo of 0.25[3].

**Supplementary Table 1:** Examination of the line profile parameters, line profile full width at half maximum (FWHM), bisector span (BIS) and Ca II H&K S-index, seeking evidence for correlations with the radial velocity (RV) measurements. For each correlation tested, the linear least squares slope with propagated uncertainty, Pearson's *r* and the *p*-value (a probability for which a small value indicates significant correlation)[2] are given. Results are shown both with and without the single S-index outlier (at -0.011; not shown in Supplementary Fig. 3) in P100.

| | FWHM - RV | | | BIS - RV | | | S-index - RV | | |
|---|---|---|---|---|---|---|---|---|---|
| Data | slope | *r* | *p*-value | slope | *r* | *p*-value | slope | *r* | *p*-value |
| P96 | 0.25 ± 0.18 | 0.19 | 0.30 | 0.15 ± 0.23 | 0.15 | 0.40 | ( 2.27 ± 2.93)×10$^2$ | 0.07 | 0.70 |
| P98 | 0.09 ± 0.11 | 0.07 | 0.53 | -0.22 ± 0.18 | -0.12 | 0.29 | (-2.70 ± 2.94)×10$^2$ | -0.17 | 0.13 |
| P100 | -0.21 ± 0.13 | -0.28 | 0.11 | 0.01 ± 0.01 | 0.09 | 0.62 | ( 0.18 ± 1.55)×10$^2$ | -0.05 | 0.79 |
| P100 clipped | | | | | | | (-6.62 ± 3.56)×10$^2$ | -0.35 | 0.04 |
| All | 0.27 ± 0.10 | 0.24 | 3.1 × 10$^{-3}$ | 0.25 ± 0.15 | 0.09 | 0.27 | ( 6.11 ± 1.52)×10$^2$ | 0.27 | 8.1 × 10$^{-4}$ |

Estimates of the Interstellar CaII H & K absorption for DMPP-1

We selected DMPP-1 as a target because it is a main sequence star with log($R'_{HK}$) = -5.16, below the basal limit of Ca II H&K line core emission, expected for even old, inactive FGK main sequence stars[4]. Consequently we conclude that the flux in the cores of the Ca II lines has been absorbed along the line of sight. Circumstantial evidence presented in the DMPP overview paper (*Haswell et al. this NA issue*) leads us to suspect this absorption arises in circumstellar material ablated from one or more close-in planet(s), but it is also possible that interstellar absorption could be responsible.

The lowest possible intrinsic value of log($R'_{HK}$) is -5.1, and the strongest interstellar absorption possible will occur if the stellar gamma velocity is exactly coincident with the radial velocity of the absorbing interstellar gas along the entire line of sight. Making these two assumptions, we can calculate the maximum expected interstellar depression of the log($R'_{HK}$) metric. We use the online interstellar log($R'_{HK}$) correction tool at http://geco.oeaw.ac.at/software.html[5] adopting the maximum column density found for lines of sight less than 60 pc[6] which is log$_{10}$N(CaII) = 11.5 cm$^{-2}$. With all of these worst-case assumptions, we obtain log $R'_{HK}$ = -5.13. We searched the literature for measurements of CaII along sightlines close to DMPP-1. The most applicable we found[6] establishes an upper limit of log$_{10}$N(CaII) < 10.2 out to 47pc, at 4 degrees on-sky separation. Using this in the online correction tool[29] we obtain log $R'_{HK}$ =-5.102.

We conclude that it is *a priori* very unlikely that the ISM is the sole cause of the log $R'_{HK}$ depression in DMPP-1. We further conclude that the *a posteriori* probability is essentially zero, given we have discovered a compact multiplanet system as predicted by DMPP's underlying hypothesis.

DMPP-1 in the context of other known compact multi-planet systems

Supplementary Figure 4a shows the DMPP-1 planets plotted in $M_p - a$ space along with the population of planets in systems hosting three or more planets with at least one planet with $a < 0.2$ AU. Bright systems with V < 8, and thus amenable to detailed follow-up are individually identified. For the Kepler systems, the plotted masses were derived from the measured radii using the relationships that also take account of the stellar incident flux[7]. We have plotted corrected mass, $M_p$, for the radial velocity planets by dividing the $M_p \sin i$ values by $<\sin i> = \pi/4$, the mean expected projection factor[8], *i.e.*, the masses are all 27.3% greater than the minimum mass values. DMPP-1 is likely to be viewed close to edge-on, so in the case of DMPP-1 the plotted masses may be overestimated.

One of the underlying scientific motivations for studying systems like DMPP-1 is to develop our understanding of planetary system formation and evolutionary processes. Two prominent demographic features of known short-period exoplanets: the Neptunian desert and the Fulton or evaporation valley[9], are both thought to be linked to the response of close-in planets to intense irradiation. Supplementary Fig. 4b shows the multiplanet systems defined above in the irradiation – planet radius plane. For the radial velocity planets, we used the masses defined above before applying the mass-radius relation[7] to place them with the more numerous Kepler objects which generally have measured radii rather than masses. DMPP-1 has the most luminous host star of the V < 8 sample. Consequently, the four DMPP-1 planets lie at the high-irradiation boundary of the observed distribution. The wedge-shaped zone of avoidance in the upper right of Supplementary Fig. 4b is the Neptunian desert.

If instead, we assume DMPP-1 is viewed exactly edge-on (*i*= 90°, $\sin i = 1$), the DMPP-1 planets have lower masses and hence radii. In this case DMPP-1c lies between the Neptunian desert and the evaporation valley, while DMPP-1d and DMPP-1e lie within the Fulton radius gap.

In either case, it is clear that DMPP-1 is a particularly informative system for further observations to test and hone our knowledges of the processes sculpting the Galaxy's population of warm and hot planets. It adds further circumstantial evidence supporting our suggestion that the circumstellar gas we infer in DMPP-1 is associated with planetary mass loss which ultimately defines an end-point in the evolution of such systems.

In Supplementary Fig. 4c, we show the orbital period – planet radius plane. The inferred radii for the DMPP-1 planets places them near the large planet boundary of the populated region. This could be because they are more massive than the typical Kepler planet. Given the uncertainty inherent in applying a single prescriptive mass-radius relationship, we will refrain from further speculation. We note if DMPP-1 and/or other DMPP systems exhibit transits, they will be extremely valuable for developing detailed quantitative tests for planetary evolution mechanisms.

## Supplementary Figures

**Supplementary Figure 1:** Left: HARPS DRS bisectors of DMPP-1 at each epoch. Right: The bisectors parallel-shifted by the measured TERRA radial velocities. Velocity sub-tic intervals are 5 m s$^{-1}$ (colour-coding for P96, P98A, P98J, P98B, P99 and P100 is respectively red, blue, magenta, green, brown, grey).

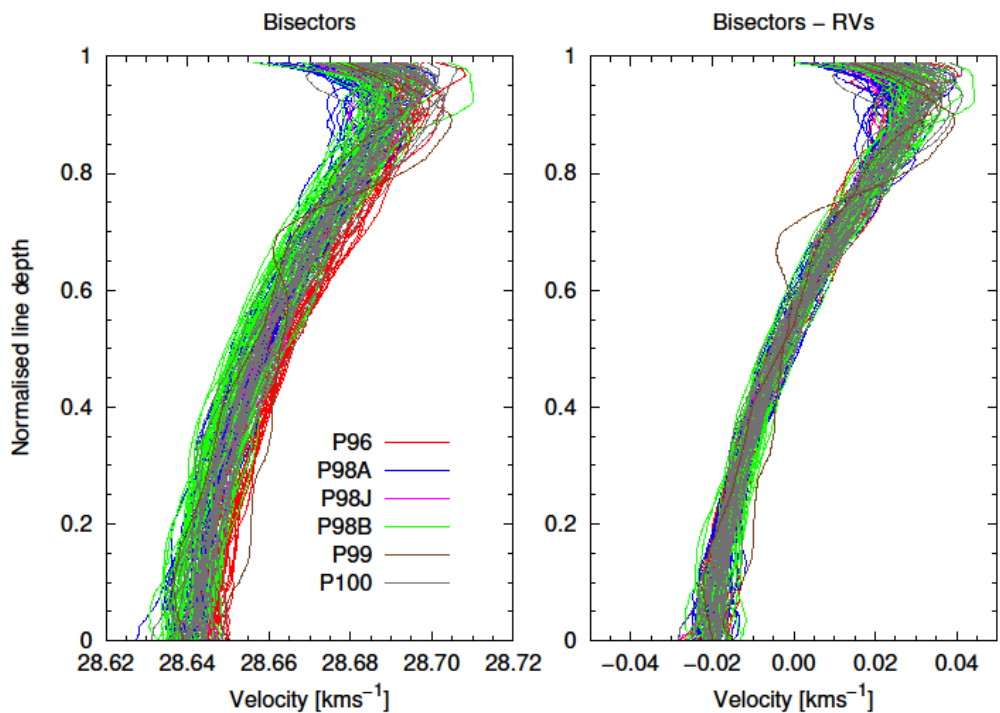

**Supplementary Figure 2:** Periodograms of the window function and Ca H&K S-index, FWHM and BIS. Vertical lines are 10, 1 and 0.1% FAPs. Vertical lines indicate the periods of the Keplerian signals identified in Fig. 2 and Table 1 (a).

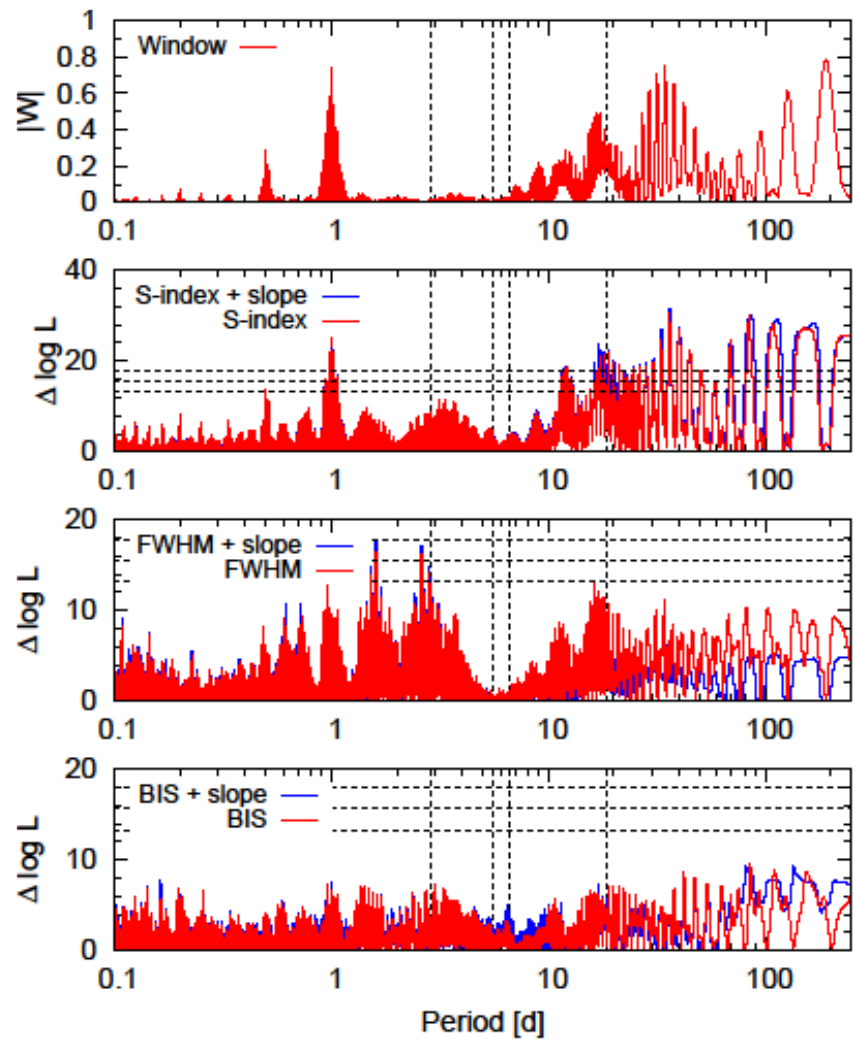

**Supplementary Figure 3:** Activity correlation plots of RV vs **(a)** FWHM, **(b)** BIS and **(c)** Ca II H&K S-index. The error bars indicate the RV 1-σ uncertainties and corresponding propagated 1-σ errors for each activity measurement.. The indices are given relative to the mean value indicated on the axis label. For S-index, the respective mean values (minus the 0.13326 global mean) at each of the 6 observing epochs, P96, P98A, P98J, P98B, P99, P100, are 0.00132, -0.00125, -0.00219, -0.00142, 0.00011 and 0.0019.

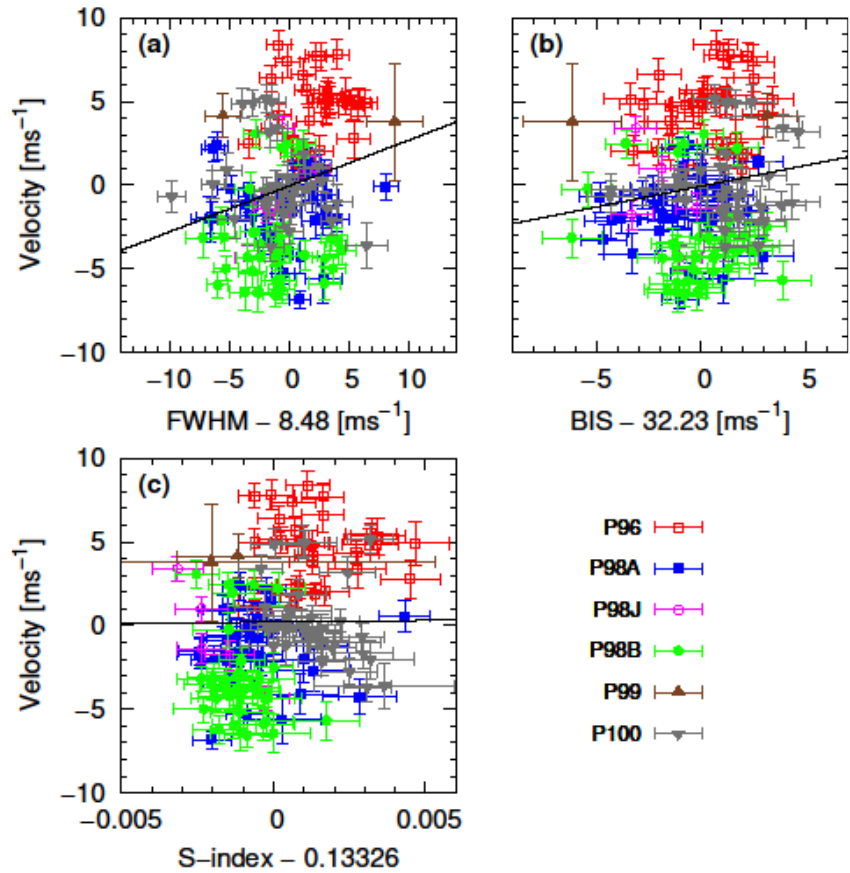

**Supplementary Figure 4:** **(a)** Orbital separation *vs* mass for DMPP-1 planets (red), other known compact multiplanet systems with V < 8 (see key), and the broader population including host stars V > 8. **(b)** Planet radius *vs* incident flux **(c)** Planet radius *vs* orbital period. For Kepler objects, the plotted masses are inferred from the measured radii[44]. For radial velocity planets, $M_p \sin i$ is divided by $<\sin i> = \pi/4$, then the corresponding radii are computed.

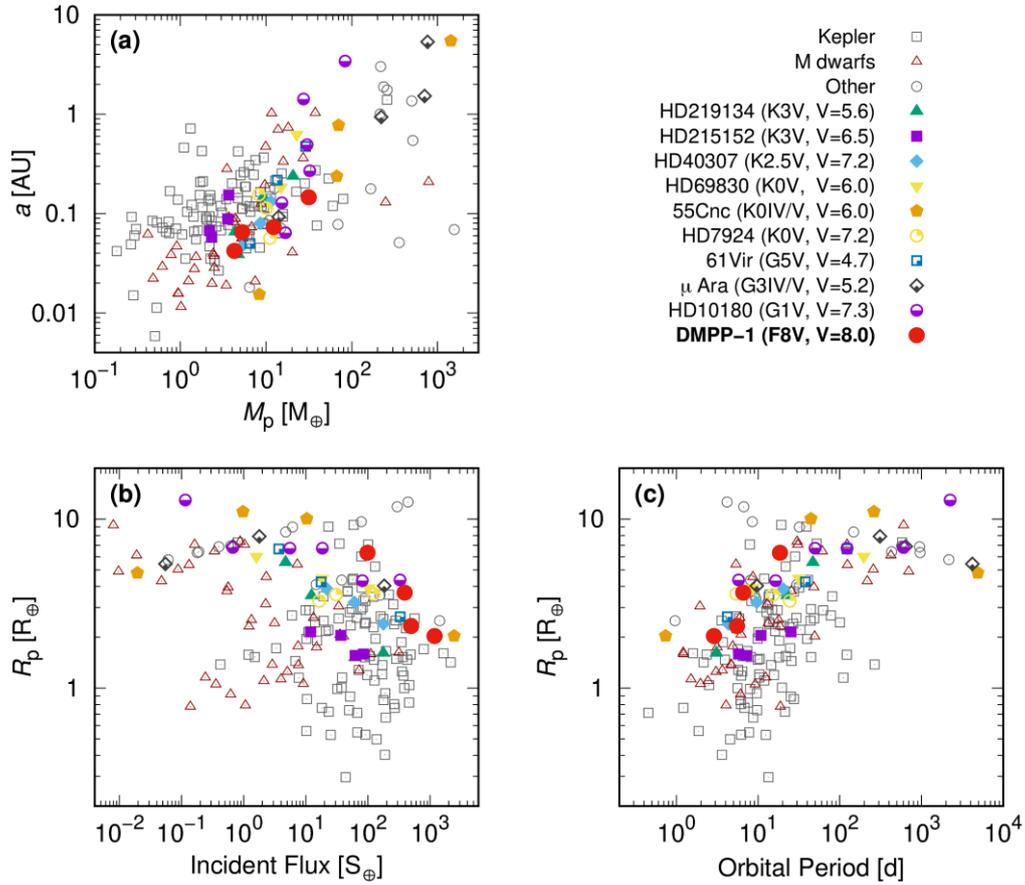